\documentclass[twocolumn,aps,prb,superscriptaddress]{revtex4-1}

\usepackage{graphicx,amsmath}
\usepackage{dcolumn}
\usepackage{bm}
\usepackage{subfigure}
\usepackage{multirow}
\usepackage{braket}

\begin{document}
\makeatletter
\def\hlinewd#1{%
\noalign{\ifnum0=`}\fi\hrule \@height #1 %
\futurelet\reserved@a\@xhline}
\makeatother

\title{Bromophenyl functionalization of carbon nanotubes : an \textit{ab initio} study.}

\author{Jason~Beaudin}
\affiliation{D\'epartement de physique et Regroupement Qu\'{e}b\'{e}cois sur les Mat\'{e}riaux de Pointe (RQMP), Universit\'e de Montr\'eal, C. P. 6128 Succursale Centre-ville, Montr\'eal (Qu\'ebec) H3C 3J7, Canada}

\author{Jonathan~Laflamme~Janssen}
\affiliation{D\'epartement de physique et Regroupement Qu\'{e}b\'{e}cois sur les Mat\'{e}riaux de Pointe (RQMP), Universit\'e de Montr\'eal, C. P. 6128 Succursale Centre-ville, Montr\'eal (Qu\'ebec) H3C 3J7, Canada}

\author{Nicholas~D.~M.~Hine}
\affiliation{Departments of Materials and Physics, Imperial College London, Exhibition Road, London SW7 2AZ, United Kingdom}

\author{Peter~D.~Haynes}
\affiliation{Departments of Materials and Physics, Imperial College London, Exhibition Road, London SW7 2AZ, United Kingdom}

\author{Michel~C\^ot\'e}
\email[Corresponding author. E-mail address: ]{michel.cote@umontreal.ca}
\affiliation{D\'epartement de physique et Regroupement Qu\'{e}b\'{e}cois sur les Mat\'{e}riaux de Pointe (RQMP), Universit\'e de Montr\'eal, C. P. 6128 Succursale Centre-ville, Montr\'eal (Qu\'ebec) H3C 3J7, Canada}

\begin{abstract}
We study the thermodynamics of bromophenyl functionalization of carbon nanotubes with respect to diameter and metallic/insulating character using density-functional theory (DFT). 
On one hand, we show that the activation energy for the grafting of a bromophenyl molecule onto a semiconducting zigzag nanotube ranges from 0.73~eV to 0.76~eV without any clear trend with respect to diameter within numerical accuracy. 
On the other hand, the binding energy of a single bromophenyl molecule shows a clear diameter dependence and ranges from 1.51~eV for a (8,0) zigzag nanotube to 0.83~eV for a (20,0) zigzag nanotube. 
This is in part explained by the transition from $sp^2$ to $sp^3$ bonding occurring to a carbon atom of a nanotube when a phenyl is grafted to it and the fact that smaller nanotubes are closer to a $sp^3$ hybridization than larger ones due to increased curvature. 
Since a second bromophenyl unit can attach without energy barrier next to an isolated grafted unit, they are assumed to exist in pairs. 
The \textit{para} configuration is found to be favored for the pairs and their binding energy decreases with increasing diameter, ranging from 4.34~eV for a (7,0) nanotube to 2.27~eV for a (29,0) nanotube. 
An analytic form for this radius dependence is derived using a tight binding hamiltonian  and first order perturbation theory. 
The $1/R^2$ dependance obtained (where $R$ is the nanotube radius) is verified by our DFT results within numerical accuracy. 
Finally, metallic nanotubes are found to be more reactive than semiconducting nanotubes, a feature that can be explained by a non-zero density of states at the Fermi level for metallic nanotubes.
\end{abstract}

\maketitle

\section{Introduction}
Since their discovery\cite{Bethune:1993ek,Iijima:1993ci}, a lot of efforts have been devoted to harnessing the exceptional characteristics of single-walled carbon nanotubes into useful electronic devices. 
Despite numerous realizations of such devices,\cite{FET1,FET2,FET3,FET4} schemes for their production at the commercial scale are scarce. 
The main problem lies in the selection and manipulation of a single nanotubes, mainly because the van der Waals force causes them to agglomerate in bundles.\cite{VDW1,VDW2} 
Moreover, the high molecular weight and hydrophobic character of carbon nanotubes makes them particularly difficult to solubilize and to handle on a substrate or in solution.\cite{aglutine1,aglutine2}

A proposed solution is to functionalize covalently the nanotube walls to modify their interaction with their environment.\cite{solution} 
A high yield method developed by J.M. Tour \textit{et al.} uses the grafting of a phenyl unit.\cite{PREMIERphenyl} 
These are commonly added to different molecules to ease their manipulation\cite{JANIE1} and increase their solubility.\cite{solution} 
It is furthermore reversible,\cite{Britz} allowing the original electronic features of the nanotubes to be retrieved by thermal annealing.\cite{JANIE2} 
Although the mechanism governing this reaction has been previously described in the literature,\cite{BLAZE} the dependence of its thermodynamics on the nanotube diameter and metallicity is not fully known and is therefore investigated in this work.

To be more specific, we report \textit{ab initio} results on the thermodynamics of the functionalization of carbon nanotubes with bromophenyl molecules. 
We study bromophenyl since it has more practical interest that phenyl, that is, because the bromine atom can be used in a successive Suzuki reaction\cite{Suzuki1,Suzuki2} to attach a vast number of other organic molecules to the grafted phenyl. 
This scheme is commonly used, for instance, in nanotube heterojunction design.\cite{SuzHete1}

This article is organized as follows~: the next section describes the computational procedure used, followed by results on the thermodynamics of the functionalization of a carbon nanotube by a single bromophenyl unit. 
Next, we assess the thermodynamics of pair formation by grafted bromophenyls. 
Finally, we derive an analytical form for the diameter dependance of the binding energy from a tight binding hamiltonian and first order perturbation theory, then verify its agreement with our computed results.

\section{Computational Details} \label{sec:comput}
All the results in this work were produced using density-functional theory (DFT)\cite{HOHENBERG,SHAM} as implemented in the ONETEP\cite{ONETEP1} code. 
The reduced computational effort for this implementation of DFT (linear scaling when applied to semiconductors) enables us to investigate the bromophenyl functionalization for a wider range of system sizes.
Also, the use of Nonorthogonal Generalized Wannier Functions (NGWFs)\cite{NGWF} ensures that the amount of empty space in the simulated system has little impact on the computation time, thus easing the simulation of large nanotubes. 
The system simulated in the present work consists of five zigzag nanotube primitive cells, with a total length of 2.13~nm and 0, 1 or 2 bromophenyl units functionalized on it. 
The chirality of the nanotubes is chosen to be zigzag since it keeps the primitive cell size minimal and allows us to simulate both semiconducting and metallic nanotubes. 
Periodic boundary conditions are applied to this system, so that the effective length of the simulated nanotube is infinite. 
Putting one single bromophenyl or bromophenyl pair per 5 zigzag nanotube primitive cells ensures that periodic images of the bromophenyl functionalization will not interact with each other. 
Therefore, this system size provides fully converged binding and activation energies, even for the metallic nanotubes. 
Due to the reduced computational effort required, we were able to investigate nanotubes of chirality up to (29,0). 
Also, the nanotubes are kept 2~nm apart from their periodic images in the direction perpendicular to the nanotube axis, to prevent any spurious nanotube-nanotube interaction.
Since the nanotubes are uncharged and have no large dipole moment, truncation of the Coulomb interaction~\cite{Hine:2011hb} is not required in this case, but could be used for other ligand species.

The electronic structure is represented by 1, 4 and 7 NGWFs with a cutoff radius of 8.0, 7.0 and 8.0 bohr for the hydrogen, carbon and bromine atoms respectively. 
An effective cutoff energy, analogous to the one in plane wave pseudopotential codes, is set to 35 Ha, which is quite sufficient for our systems (c.f. E.R. Margine and \textit{al}\cite{BLAZE}). 
The PBE\cite{PBE} approximation for the exchange-correlation energy is used throughout this work. 
Spin-polarized calculations are used whenever the system contains an odd number of electrons.
A level of convergence was achieved such that the calculated activation and binding energies have a numerical accuracy of 50~meV per primitive cell (consisting of 5 zigzag nanotube primitive cells). 
Geometric relaxations of each structure were then carried out with a tighter convergence criteria of 10 meV per primitive cell.
The optimization of the NGWFs within ONETEP results in a smooth potential energy surface from which accurate ionic forces can be calculated \cite{forces} and
eliminates the basis set superposition error\cite{onetepbsse} present in traditional DFT implementations based on localized basis sets.
This feature allows convergence criteria as tight as the one above to be reached.

\section{Activation energy}
\subsection{Procedure}
\begin{figure}
\centering
\subfigure[ ~Initial configuration]{
\includegraphics[width=0.75 \linewidth,angle=90,viewport=200 100 1050 485,clip=true]{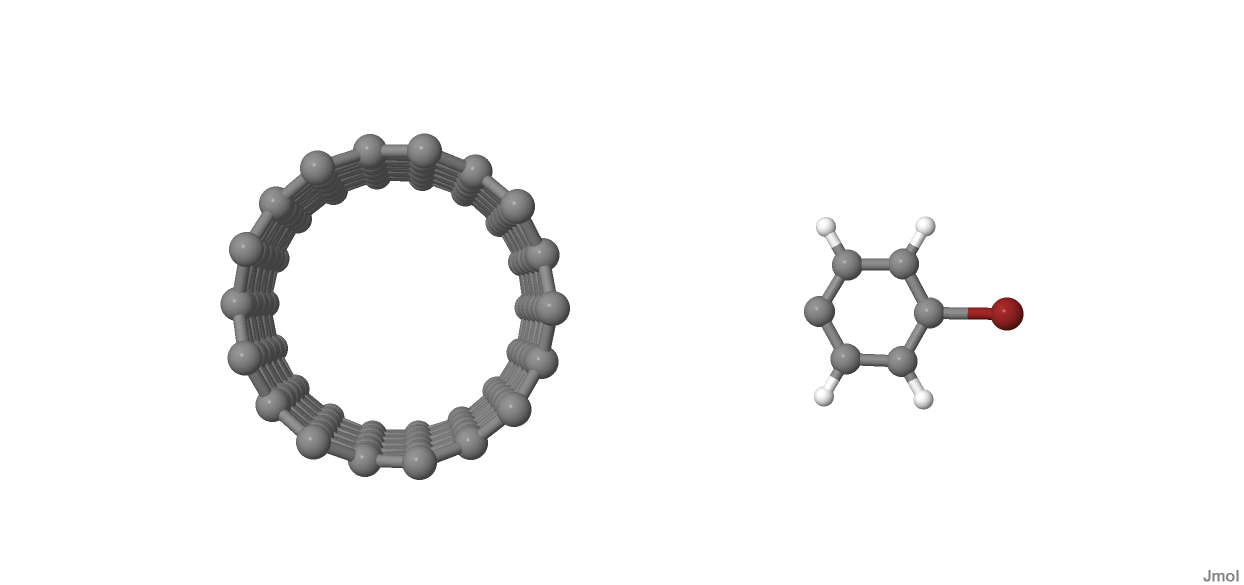}
\label{fig:departbarriere}}
\subfigure[ ~Final configuration]{
\includegraphics[width=0.75 \linewidth,angle=90,viewport=200 50 1050 535,clip=true]{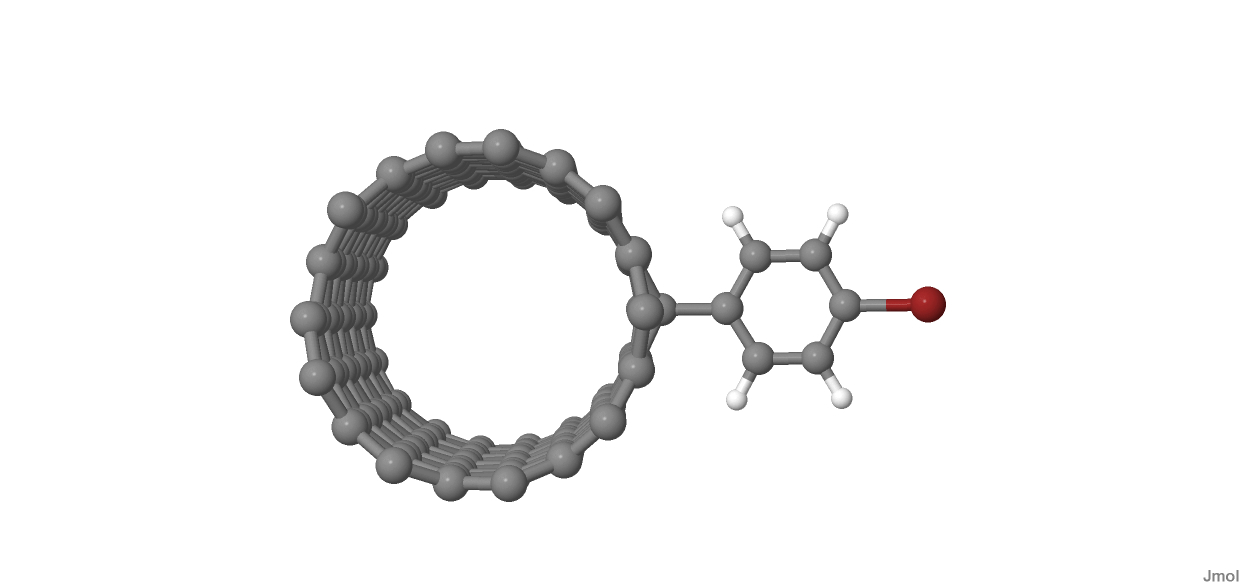}
\label{fig:finbarriere}}
\caption[Extremum scheme constituting the reaction barrier of a bromophenyl unit with a carbon nanotube.]
{\label{fig:departfin}
Initial and final configurations used to simulate the bromophenyl functionalization of a (9,0) carbon nanotube. 
The distance between the nanotube and the bromophenyl unit is set to 7~\AA\; in the initial configuration to prevent any interaction between the two systems. 
In the final configuration, the relaxed bromophenyl-nanotube bond length is 1.54 \AA.}
\end{figure}

In this section, we report our calculated activation and binding energies for the functionalization of a nanotube with one bromophenyl unit. 
In both cases, the first step is to relax a system where the pristine nanotube and the bromophenyl unit are 7~\AA\;apart (see Fig.~\ref{fig:departbarriere}) to find the ground state energy of the isolated reactants. 
Then, a functionalized nanotube (see Fig.~\ref{fig:finbarriere}) is relaxed to obtain the ground state energy of the product. 
The binding energy is then obtained by taking the difference between the two results.

\begin{figure}
\centering
\includegraphics[width=1.0 \linewidth]{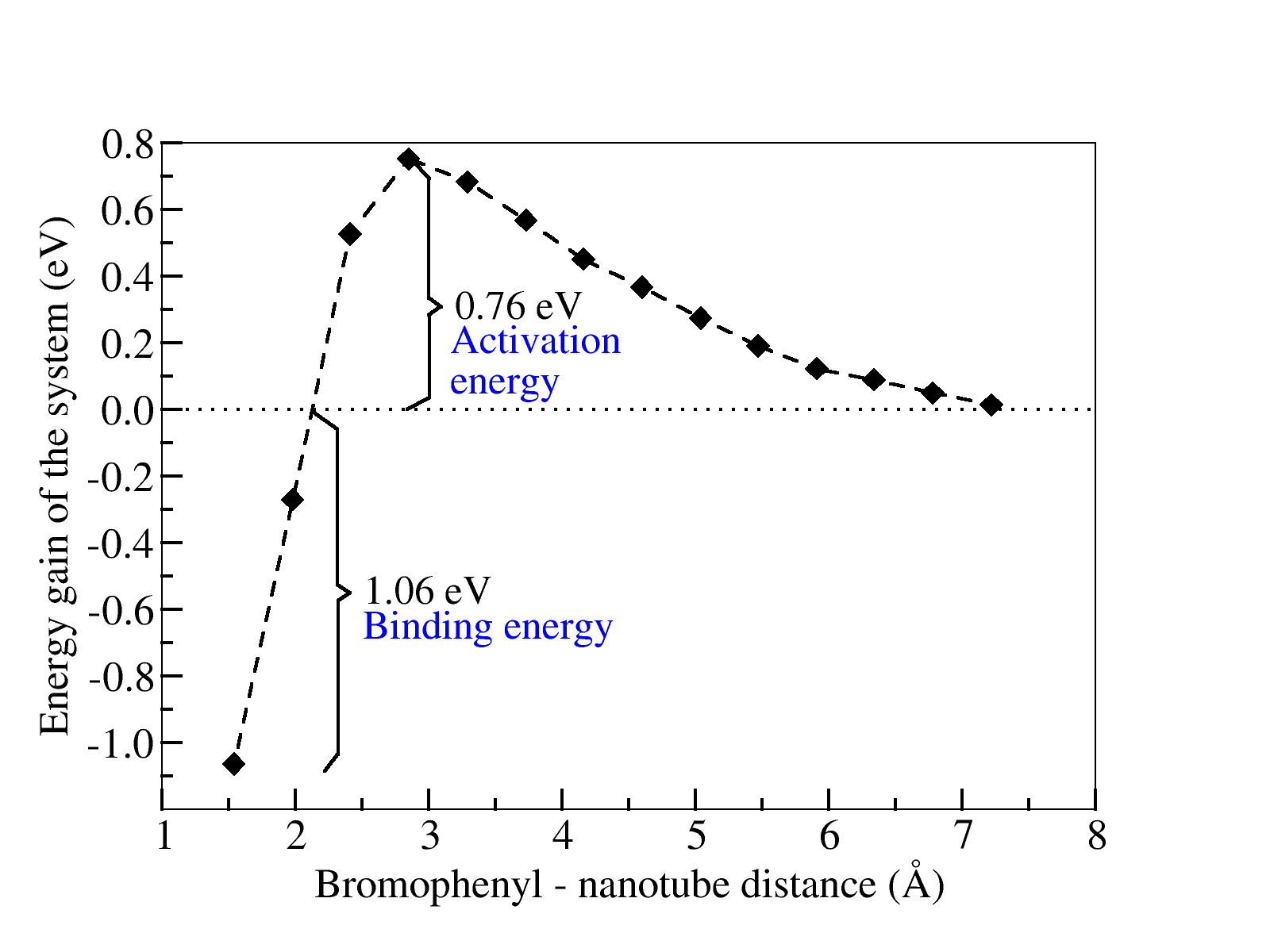}
\caption{
\label{fig:barriere13x0}
Reaction barrier for the functionalization of the (13,0) nanotube by a bromophenyl unit. 
The dashed line is a guide for the eye and is set to the ground state energy of the reactants. 
Each point between the initial and final ones is a linear interpolation between the two geometries shown in Fig.~\ref{fig:departfin}.}
\end{figure}

To find the activation energy, we linearly interpolate the atomic positions of 12 intermediate systems between the isolated reactants and the final product. 
The ground state energy of each interpolated system is then calculated without any geometric relaxation. 
The top of the energy barrier is taken to be the highest energy value obtained among the 12 interpolated calculations. 
The activation energy is obtained by taking the difference between the top of the energy barrier and the ground state energy of the isolated reactants. 
As an example, a graph of the total energy with respect to bromophenyl-nanotube distance is given for a (13,0) nanotube in Fig.~\ref{fig:barriere13x0}. 
We obtain an activation energy of~0.76 eV and a binding energy of 1.06~eV which are of the right order of magnitude according to previous studies.\cite{BLAZE}

It should be noted that the activation energies obtained here are in fact upper bounds to the exact DFT ones. 
Rigorously, a complete relaxation for each geometric distance between the nanotube and the bromophenyl unit should have been performed. 
This would have lowered slightly our estimation of the activation energy and greatly increased the computational cost. 
Still, this should not prevent diameter or metallicity dependence from being observed in our calculations. 
However, our activation energy calculations might overestimate by more than 50~meV the exact DFT prediction, despite the tight numerical precision of our calculations.

To get an intuitive understanding of the activation and binding energies trends with respect to nanotube diameter and metallicity, we divided both kind of energies into three parts : the contribution from the deformation of the nanotube, the contribution from the deformation of the bromophenyl unit and the remainder, labeled the electronic contribution. 
For the binding energy, the nanotube deformation contribution is obtained from the difference between the total energy of a system having the same geometry as the functionalized nanotube, but from which the bromophenyl has been removed, and the total energy of the pristine nanotube. 
The same scheme was applied for the activation energy, except that the functionalized nanotube is replaced with the system configuration at the top of the energy barrier. 
The bromophenyl deformation contribution is obtained with a scheme analogous to the nanotube deformation contribution.

\subsection{Results}
\begin{table}
\caption{\label{decomp_energie_top}Contributions to activation energies (in eV) for a bromophenyl unit functionalization.}
\begin{tabular}{c|c|c|c|c|}
\cline{2-5}
& \multicolumn{3}{c|}{Semiconducting} & Metallic \\ \hline
\multicolumn{1}{|l|}{\textbf{Nanotube}} & \textbf{(8,0)} & \textbf{(13,0)} & \textbf{(20,0)} & \multicolumn{1}{|c|}{\textbf{(9,0)}} \\ \hlinewd{.5mm}
\multicolumn{1}{|l|}{Nanotube deformation} & 0.70 & 0.81 & 0.86 & \multicolumn{1}{|c|}{0.63}  \\ \hline
\multicolumn{1}{|l|}{Bromophenyl deformation} & 0.06 & 0.06 & 0.06 & \multicolumn{1}{|c|}{0.10}  \\ \hline
\multicolumn{1}{|l|}{Electronic contribution} & -0.03 & -0.11 & -0.15 & \multicolumn{1}{|c|}{-0.09}  \\ \hlinewd{.5mm}
\multicolumn{1}{|l|}{Activation energy} & 0.73 & 0.76 & 0.77 & \multicolumn{1}{|c|}{0.64} \\ \hline
\end{tabular}
\end{table}

\begin{table}
\caption{\label{decomp_energie_bas}Contributions to binding energies (in eV) for a bromophenyl unit functionalization.}
\begin{tabular}{c|c|c|c|c|}
\cline{2-5}
& \multicolumn{3}{c|}{Semiconducting} & Metallic \\ \hline
\multicolumn{1}{|l|}{\textbf{Nanotube}} & \textbf{(8,0)} & \textbf{(13,0)} & \textbf{(20,0)} & \multicolumn{1}{|c|}{\textbf{(9,0)}}  \\ \hlinewd{.5mm}
\multicolumn{1}{|l|}{Nanotube deformation} & 1.12 & 1.30 & 1.38 & \multicolumn{1}{|c|}{1.17}  \\ \hline
\multicolumn{1}{|l|}{Bromophenyl deformation} & 0.11 & 0.12 & 0.11 & \multicolumn{1}{|c|}{0.12}  \\ \hline
\multicolumn{1}{|l|}{Electronic contribution} & -2.74 & -2.47 & -2.33 & \multicolumn{1}{|c|}{-3.14}  \\ \hlinewd{.5mm}
\multicolumn{1}{|l|}{Binding energy (absolute value)} & 1.51 & 1.05 & 0.84 & \multicolumn{1}{|c|}{1.85}  \\ \hline
\end{tabular}
\end{table}

Results for the activation energy are presented in Table~\ref{decomp_energie_top}. 
For the semiconducting nanotubes, the deformation contribution increases with their diameter. 
However, the electronic contribution has the opposite trend, so that the two effects cancel out up to numerical precision. 
Thus, the activation energy is found to be between 0.73 and 0.76~eV, without any significant diameter dependence. 
However, the metallic nanotube (9,0), with an activation energy of 0.64~eV, shows a reactivity significantly higher than the semiconducting nanotubes. 
This trend agrees with other DFT studies on $\text{NO}_2$ \cite{SEO} and carbene \cite{BETTINGER} nanotube functionalization and is associated with a non-zero electronic density of states (DOS) at the Fermi level of metallic nanotubes.\cite{SEO} 
As we shall see later, all metallic nanotubes in the present study are more reactive than semiconducting tubes of similar diameter.

Results for the binding energy are presented in Table~\ref{decomp_energie_bas}. 
For the semiconducting nanotubes, the deformation energy cost increases with diameter. 
This is easy to understand intuitively. 
The bromophenyl functionalization involves changing the hybridization of a carbon atom of the sidewall of the nanotube from $sp^2$-like to $sp^3$-like.\cite{YSLEE} 
For larger diameters, the geometry of the wall is close to planarity and to ideal geometry of $sp^2$ bonding. 
Thus, when the nanotube's diameter increases, it becomes more energetically unfavorable to undergo geometrical deformation to $sp^3$ bonding. 
That explains why the deformation contribution to the binding energy favors functionalization of small nanotubes over large ones. 
This analysis applies both to binding and activation energies, thus explaining why deformation energy favors small tube functionalization in both cases.

The electronic contribution to the binding energy also favors small nanotubes. 
Since the bromophenyl deformation energy shows no trend with respect to diameter and both other contributions favor small nanotubes, the total binding energy is stronger when the nanotube is smaller.

The binding energy trend with respect to metallicity shows, for the same reason given in the activation energy discussion, that the electronic contribution is higher for the metallic nanotubes than for the semiconducting ones. 
Since both deformation contributions to the binding energy remain similar for nanotubes of similar size, whether they are semiconducting or metallic, the overall result is that metallic nanotubes bond more strongly with bromophenyl than semiconducting ones.

In summary, our results show that the activation energy of bromophenyl functionalization of semiconducting nanotubes has almost no diameter dependence, although the binding energy is significantly higher for small tubes, making their functionalization more stable. 
This is in part explained by the fact that  nanotubes with higher curvature are closer to a $sp^3$ hybridization. 
Also, thanks to the presence of a non-zero DOS at their Fermi level, metallic nanotubes can be functionalized by bromophenyl units with more ease and more stability than semiconducting nanotubes of similar diameter. 
This could provide an easy way to sort semiconducting nanotubes from metallic ones.

\section{Functionalization in pairs}
\subsection{Procedure}
The study presented in the previous section explores how a bromophenyl unit reacts with a pristine nanotube. 
However, it has been shown in the literature that, at room temperature, a single bromophenyl grafted onto a nanotube can desorb or migrate on its surface.\cite{BLAZE} 
Therefore, a stable bromophenyl unit on a nanotube cannot be isolated as pictured in Fig.~\ref{fig:departfin}. 
It is believed that it must instead be adjacent to another bromophenyl unit.\cite{BLAZE} 
The motivation for this hypothesis can be explained qualitatively and  was introduced by the groups of C.~Dyke\cite{DYKE} and G.~Schmidt.\cite{SCHMIDT}
Just before grafting onto the nanotube, the bromophenyl radical gets an unpaired electron from the separation of phenyldiazonium ($R\mathrm{C}_6\mathrm{H}_4\mathrm{N}_2^+$) and is therefore highly reactive.

\begin{figure}
\centering
\includegraphics[width=1.00 \linewidth]{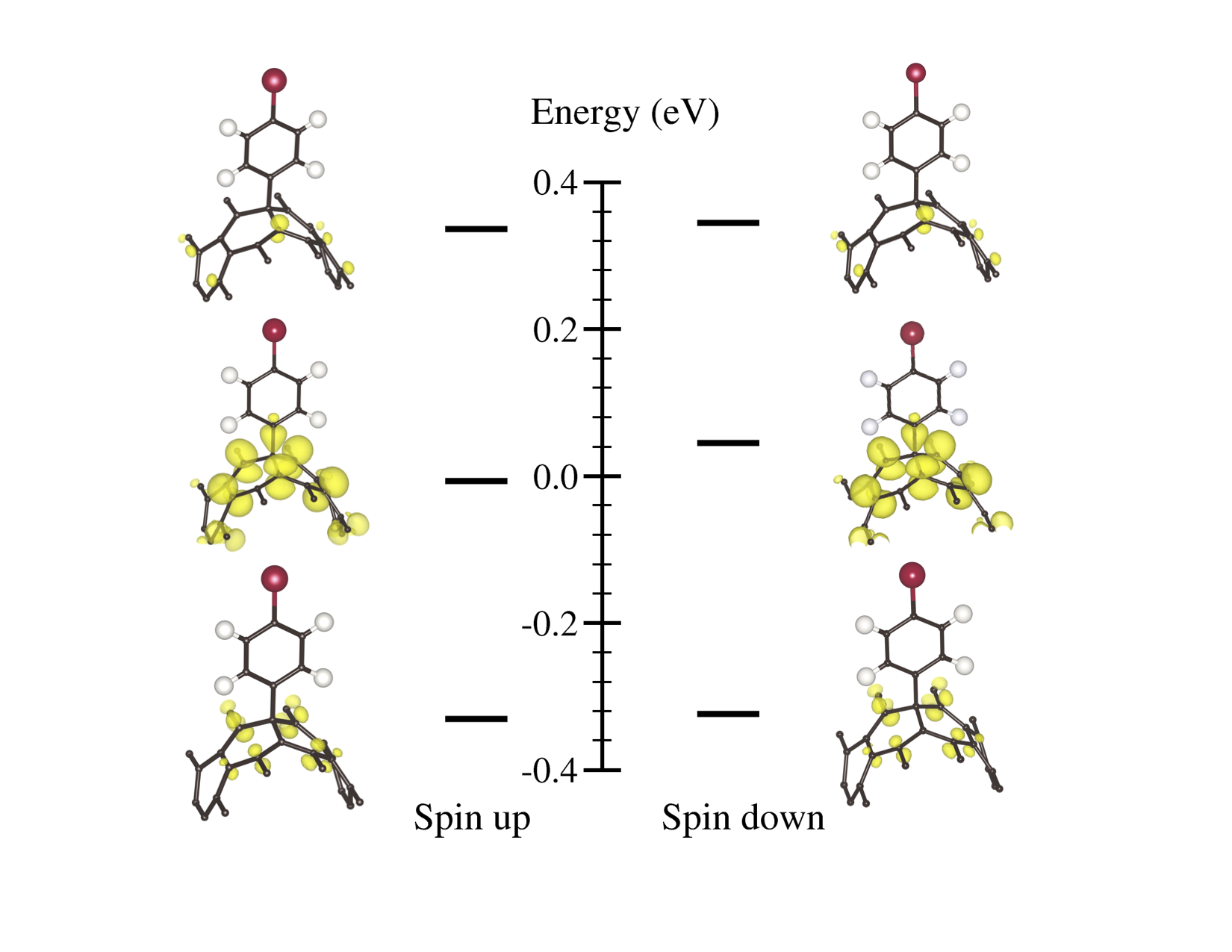}
\caption{
\label{fig:nivenerg}
Energy levels and their corresponding electronic densities for orbitals near the Fermi level of functionalized (13,0) nanotube. 
Only the functionalized site and the surrounding atoms are shown. }
\end{figure}

When the first bromophenyl unit covalently bonds, its half-filled orbital mixes with the orbitals of the tube. 
This adds two new electronic states in the gap of the pristine nanotube according to our calculations. 
One is associated with the filled state of the half-filled orbital of the bromophenyl unit and constitutes the highest occupied molecular orbital (HOMO) of the functionalized nanotube. 
The other electronic state is associated with the empty state of the half-filled orbital of the bromophenyl unit and constitutes the lowest unoccupied molecular orbital (LUMO) of the functionalized tube. 
Fig.~\ref{fig:nivenerg} shows the electronic density of the orbitals near the Fermi level of the functionalized (13,0) nanotube. 
Only a section of the sidewall of the tube close to the functionalized site is shown to ease visualization. 
We notice that the LUMO has exactly the same shape as the HOMO state and is 50 meV above the Fermi level. 
Both the HOMO and LUMO states are well localized near the functionalized site, whereas the other orbitals are spread over the nanotube. 
The other nearest states are further apart in energy from the Fermi level, about 350 meV from it. 
Since the new LUMO only exists a few angstroms around the first bromophenyl unit on the nanotube's sidewall, the functionalization of the second bromophenyl, eased by the higher electronic affinity around the first one, takes place close to it. 
Thus, at room temperature, bromophenyls should be found in pairs on the surface of a functionalized nanotube.

\begin{figure}
\centering
\includegraphics[width=0.75 \linewidth,viewport=230 310 510 570,clip=true]{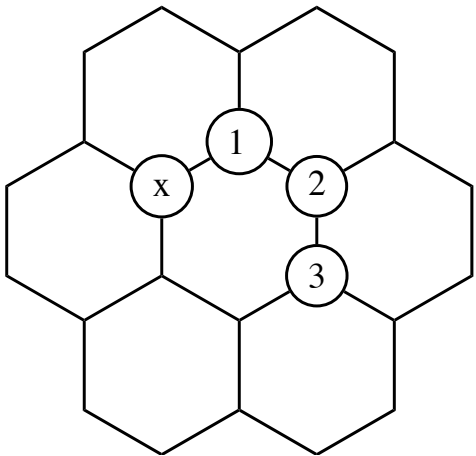}
\caption[Different configurations for a pair of bromophenyl.]{
\label{fig:sitereaction}
Three possible configurations for a bromophenyl pair. 
The "x" shows the first functionalized site. 
The location of the second unit gives the \textit{ortho} (1), \textit{meta} (2) and \textit{para} (3) configurations.}
\end{figure}

Since we are interested in the thermodynamics of the functionalization, it is relevant to investigate the stability of the final product obtained, that is, the binding energy of pairs of bromophenyls on carbon nanotubes.
Yet, a calculation of the binding energy of the pairs requires a precise knowledge of the configuration of these pairs. 
Since the LUMO is strongly localized around the first bromophenyl unit, we expect the second bromophenyl to bind more strongly closer to the first one. 
We then have three likely configurations, pictured in Fig.~\ref{fig:sitereaction} and named \textit{ortho}(first adjacent atom), \textit{meta}(second adjacent atom) and \textit{para}(third adjacent atom).
To evaluate the likeliness of the different reaction sites, let's first neglect the effect of the sidewall curvature of the nanotube. 

It has been shown in the literature, using \textit{ab initio} calculations, that the \textit{meta} configuration of aryl groups on a graphene sheet is unstable\cite{JIANG} in accordance with the aryl addition rule. 
Moreover, the LUMO state illustrated in Fig.~\ref{fig:nivenerg} is localized on the \textit{ortho} and \textit{para} sites, but not the \textit{meta} sites.
The functionalization of the former ones is therefore favored over the latter. 

However, steric constraints will lower the binding energy of the second phenyl if it binds on atoms adjacent to the first reaction site, thus disfavoring the \textit{ortho} site.  

In addition, the binding energies of phenyls on graphene for these configurations were found by Margine \textit{et al.} to be 1.25 eV for the \textit{ortho} site and 1.51 eV for the \textit{para} site.\cite{BLAZE} 
The higher binding energy of the \textit{para} configuration over the \textit{ortho} configuration shows that although the \textit{ortho} configuration has larger LUMO electronic density than the \textit{para} configuration, the steric constraints disfavor the closest site.
 
In our study, a hydrogen of the phenyl is replaced by a bromine atom, which should further increase the steric constraints. 
Indeed, our results also show that \textit{para} configuration is favored over the \textit{ortho} configuration in our systems. 
Therefore, we limit our study to the former case. 

The reader should note that no reaction barriers are shown for the second functionalization. 
First, it is very difficult to define the reaction pathway at this stage. 
Second, the activation energy for the second bromophenyl is known to be smaller than for the first reaction\cite{BLAZE} and thus does not limit the reaction rate.

If we now take into account the curvature of the sidewall, we find that there are two possible variants for the \textit{para} configuration on a zigzag nanotube. 
The imaginary line linking the two grafted sites can either be parallel to the axis of the tube or make an angle of $60^{\circ}$ with it. 
Both cases should show different binding energies since the curvature of the nanotube affects their geometry. 
Therefore, for pairs of bromophenyls in the \textit{para} configuration, we investigate the magnitude of this difference in addition to the diameter and metallicity dependence of their binding energy.

\subsection{Results}
\begin{figure}
\centering
\includegraphics[width=1.00 \linewidth,viewport=14 0 530 430,clip=true]{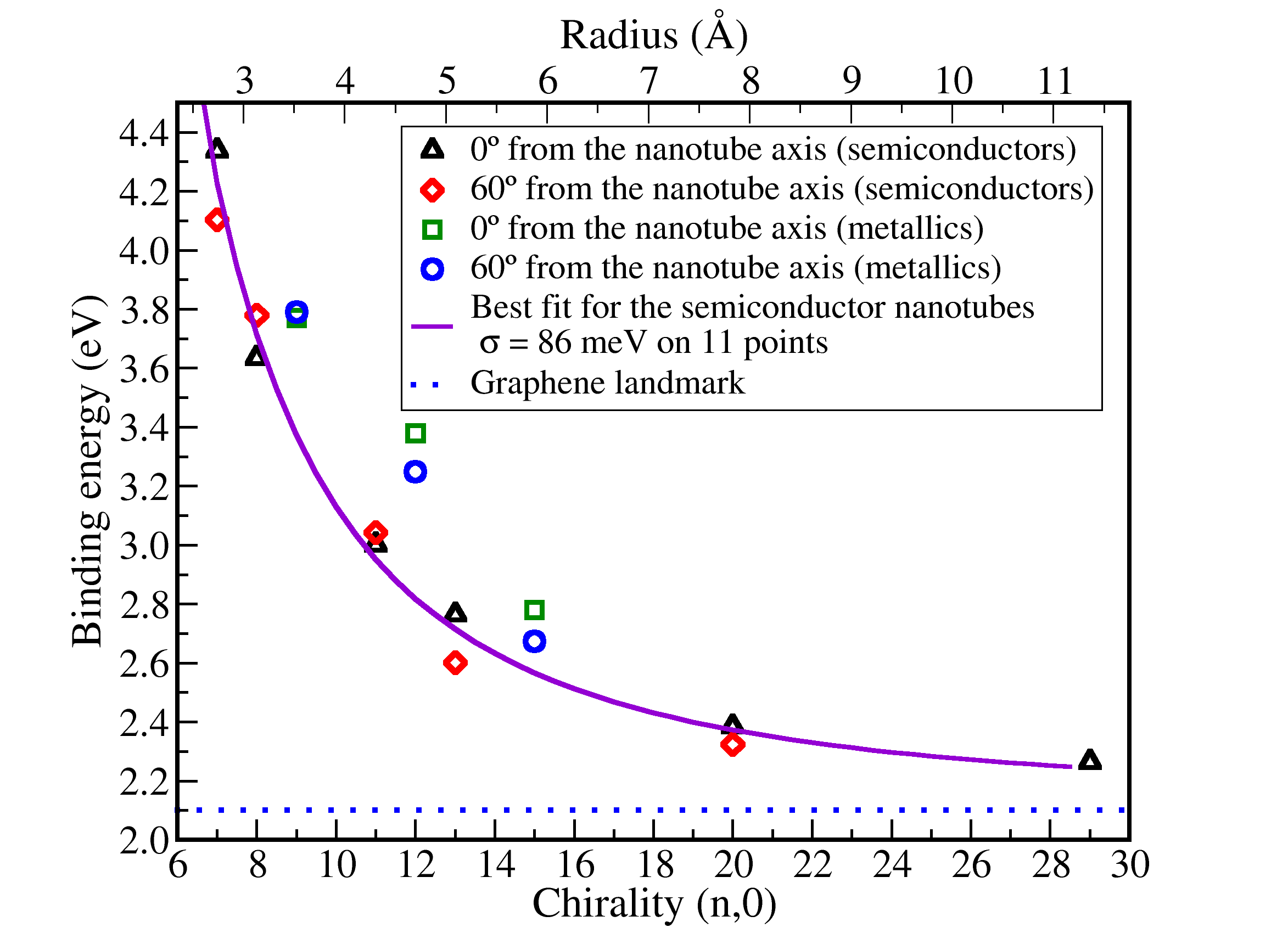}
\caption[The graph]{
\label{fig:thegraph}
Binding energy of bromophenyl pairs on zigzag nanotubes for two \textit{para} variants ($0^\circ{}$ and $60^\circ{}$ relative to the tube axis). 
The continuous line shows the best fit of Eq.~\ref{eq:tendance} on the semiconducting nanotubes data. 
We obtain $E_{\text{binding}}=15.00 \text{~eV~\AA}^2 / R^2 +2.11~\text{eV}$. 
The standard deviation $\sigma=86$~meV  is similar to our numerical accuracy of 50~meV, which supports the validity of our model (Eq.~\ref{eq:tendance}).}
\end{figure}

The results are shown in Fig.~\ref{fig:thegraph}. 
The difference of binding energy between the two variants of \textit{para} configurations ($0^\circ{}$ and $60^\circ{}$) does not show any clear trend with respect to diameter or metallicity and its magnitude remains small in all cases. 
Therefore, we cannot conclude whether there is a variant favored in the functionalization process.

However, the binding energy trend with respect to diameter is clear~: bromophenyl pairs are more stable on small nanotubes. 
This conclusion agrees with previous \textit{ab initio} studies of nanotube functionalization with phenyls\cite{BLAZE} and $\text{NO}_2$.\cite{SEO} The fit in Fig.~\ref{fig:thegraph} is explained in the next section.

\begin{table}
\caption{\label{comparaisonpaire}Binding energies (in eV) for bromophenyl pairs (\textit{para} configuration, $0^{\circ}$ relative to the axis, see Fig.~\ref{fig:thegraph}) and for two isolated units.}
\begin{tabular}{c|c|c|c|c|}
\cline{2-5}
& \multicolumn{3}{c|}{Semiconducting} &  Metallic \\ \hline
\multicolumn{1}{|p{4.0cm}|}{\textbf{Nanotube}} & \textbf{(8,0)} & \textbf{(13,0)} & \textbf{(20,0)} & \multicolumn{1}{|c|}{\textbf{(9,0)}} \\ \hlinewd{.5mm}
\multicolumn{1}{|p{4.0cm}|}{\raggedright Bromophenyl pair} & 3.63 & 2.76 & 2.38& \multicolumn{1}{|c|}{3.77}  \\ \hline
\multicolumn{1}{|l|}{\raggedright Single bromophenyl unit $\times$ 2} & 3.02 & 2.12 & 1.66 & \multicolumn{1}{|c|}{3.72} \\ \hline
\end{tabular}
\end{table}

Table~\ref{comparaisonpaire} compares the binding energy of \textit{para} paired bromophenyls units and isolated units. 
The results confirm the pair configuration to be more stable than the isolated configuration for all semiconducting cases studied. 
However, they cast doubt in the case of metallic nanotubes since the pairs are only 50 meV more stable than isolated units and could therefore separate easily at room temperature ($k_BT=25$ meV). 
This observation is in agreement with the explanation given earlier for the pairing of bromophenyls. 
Indeed, since the pairing occurs because the LUMO of a nanotube functionalized with a single bromophenyl is well inside the electronic gap of the pristine nanotube, the proposed mechanism cannot account for an eventual pairing of the bromophenyls on metallic nanotubes since they have no gap. 

\section{Theoretical background}
It is interesting to derive an approximate expression for the diameter dependance of the binding energy of a bromophenyl pair on a nanotube. 
At first order in perturbation theory, it is given by the expectation value of the Hamiltonian of the bromophenyl pair, $H'$, in the eigenstate of the pristine nanotube (most) involved in the bond formation, $\mid\pi_n\rangle$. 
Since the other eigenstates have a smaller overlap with the orbitals of the phenyl graft, we do not expect them to contribute significantly to the bromophenyl-nanotube bond. 

With a tight binding approach, following the notation of M. Pudlak \textit{et al.},\cite{PUDLAK} the $\ket{\pi}$ orbital of a nanotube can be written as:
\begin{equation}
\begin{split}
\ket{\pi_n} = & \tan\theta \sqrt{\frac{3\cos2\theta-1}{3\cos2\theta}}\ket{s} \\
 &+ \frac{\tan\theta}{\sqrt{3\cos2\theta}}\mid p_y \rangle + \frac{\sqrt{\cos2\theta}}{\cos\theta}\mid p_z \rangle,
\end{split}
\end{equation}
where $\theta$ is a measure of the curvature and is defined as:
\begin{equation}
\sin\theta=\frac{\sqrt{3}a}{4R},
\end{equation}
where $a=1.42$ \AA~is the length of a covalent bond in the nanotube and R is the radius. 
To first order in static non-degenerate perturbation theory, we have:
\begin{equation}
\braket{ s | H' | p_y} =\braket{ s | H' | p_z} = \braket{ p_y | H' | p_z} = 0.
\end{equation}
Therefore,
\begin{equation}
\begin{split}
\braket{ \pi_n | H' | \pi_n} = & \tan^2\theta \Big(\frac{3\cos2\theta-1}{3\cos2\theta}\Big) \braket{ s | H' | s } \\
&+ \frac{\tan^2\theta}{3\cos2\theta} \braket{ p_y | H' | p_y } \\
& + \frac{\cos2\theta}{\cos^2\theta} \braket{ p_z | H' | p_z }.
\end{split}
\end{equation}
With a Taylor expansion and some algebra, we find that for small values of 1/R ($R \gg a$), at leading order~:
\begin{equation}
\begin{split}
\braket{ \pi_n | H' | \pi_n} = & \braket{ p_z | H' | p_z } + \left( \frac{2}{3} \braket{ s | H' | s } \right. \\
& \left. + \frac{1}{3} \braket{ p_y | H' | p_y } - \braket{ p_z | H' | p_z } \right) \frac{3a^2}{16R^2} .
\end{split}
\end{equation}

Therefore, the trend with respect to diameter of the binding energy for a bromophenyl pair functionalization can be written as:
\begin{equation}
\label{eq:tendance}
E_{\text{binding}} = \alpha \frac{1}{R^2}+c,
\end{equation}
where $c$ is the binding energy for graphene. 
This argument also applies to the binding energy of a single bromophenyl. 
However, in the latter case, we do not have enough results to assess the accuracy of Eq.~\ref{eq:tendance}.

A fit of Eq.~\ref{eq:tendance} on binding energies for pairs of bromophenyl on semiconducting nanotubes is shown in Fig.~\ref{fig:thegraph}. 
The agreement is excellent. 
We find:
\begin{equation}
\label{eq:resultpaire}
E_{\text{binding}} =15.00 \text{~eV} \frac{\text{\AA}^2}{R^2}+2.11 \;\;\text{eV} ,
\end{equation}
with a standard deviation of $\sigma=86$~meV. 
However, for metallic nanotubes, the amount of available data was insufficient for a fit of Eq.~\ref{eq:tendance} to be meaningful. 

The discrepancy between the fit on semiconducting nanotube data (Eq.~\ref{eq:resultpaire}) and our results is similar to our numerical accuracy ($\sigma_{\mathrm{fit}}=86$~meV $\sim \sigma_{\mathrm{convergence}}\approx 50$~meV). 
Also, our fitted value of $c$ is 2.11 eV, in very close agreement to the calculated binding energy of 2.10~eV, which is not part of the fitted data.
Therefore, Eq.~\ref{eq:tendance} accurately describes the diameter dependance of semiconducting nanotubes functionalized with bromophenyl pairs.
Since no properties exclusive to bromophenyl has been assumed in the derivation, it should accurately describe any covalent functionalization of nanotubes where first order perturbation theory remains valid, i.e.\ where the binding energy is of the same order of magnitude as the cases studied here.

The binding energy also shows a clear trend with metallicity~: it is significantly higher for metallic nanotubes than for semiconducting nanotubes of similar diameter. 
This is expected since the explanation given for the activation energy trend with metallicity remains valid for the binding energy of pairs. 
Together with the activation energies, these results confirm that bromophenyl functionalization thermodynamically favors nanotubes of smaller diameter and metallic nanotubes over semiconducting ones.

\section{Conclusion}

Our activation and binding energy calculations show that bromophenyl functionalization is selective with respect to nanotube size and metallicity~: functionalization of smaller and/or metallic nanotubes is thermodynamically favored. 
However, this selectivity is stronger for smaller nanotubes while the experimental tubes are relatively large. 
Therefore, it may be challenging to observe this selectivity in experimental samples, due to others factors becoming of equal importance as compared to the diameter, such as the family of the nanotube.

\section{Acknowledgments}

We would like to thank NSERC, RQMP and Thomas Young Centre for funding.
We would also like to thank Calcul Quebec and Imperial College's High Performance Computing Service for computational resources.
PHD would like to acknowledge support of the EPSRC (UK) under grant no.\ EP/G05567X/1 and the Royal Society University Research Fellowship.

\bibliography{references}

\end{document}